%% file: paper.tex
\documentclass[aps,preprintnumbers,superscriptaddress,nofootinbib,aps,prd,twocolumn,floatfix]{revtex4-1}
\pdfoutput=1

\usepackage{multirow,array}
\usepackage{amsmath,amssymb}
\usepackage{graphicx}
\usepackage{color}
\usepackage{url}
\usepackage{slashed}
\usepackage{multirow}
\usepackage{footmisc}
\usepackage{hyperref}
\usepackage{braket}
\usepackage{colortbl}

\hypersetup{
  pdfauthor={Dorival Goncalves, Tao Han, Satyanarayan Mukhopadhyay} 
  pdftitle={Off-shell Higgs Probe to Naturalness},
  pdfkeywords={QCD, NLO, Higgs Physics}
}

\input{declare}

\begin{document}
\title{Off-shell Higgs Probe to Naturalness}

\author{Dorival Gon\c{c}alves} 
\author{Tao Han}
\author{Satyanarayan Mukhopadhyay} 
\affiliation{PITT PACC, Department of Physics and Astronomy, University of Pittsburgh, 3941 O'Hara St., Pittsburgh, PA 15260, USA} 

\preprint{PITT-PACC-1711}

\begin{abstract}
 \noindent
Examining the Higgs sector at high energy scales through off-shell Higgs production can potentially shed light on the naturalness problem of the Higgs mass. We propose such a study at the LHC by utilizing a representative model with a new scalar field ($S$) coupled to the Standard Model Higgs doublet ($H$) in a form $ |S|^2 |H|^2$. In the process $p p \rightarrow h^* \rightarrow ZZ$, the dominant momentum-dependent part of the one-loop scalar singlet corrections, especially above the new threshold at $2m_S$, leads to a measurable deviation in the differential distribution of the $Z$-pair invariant mass, in accordance with the quadratic divergence cancellation to the Higgs mass. We find that it is conceivable to probe such new physics at the $5\sigma$ level at the high-luminosity LHC, improving further with the upgraded $27$~TeV LHC, without requiring the precise measurement of the Higgs boson total width. The discovery of such a Higgs portal could also have important implications for thermal dark matter as well as for electroweak baryogenesis.

\end{abstract}

\maketitle

\noindent
{\bf I. Introduction}\\  
Off-shell production of the Higgs boson has substantial event rate at the LHC~\cite{offshell,offshell_ex}. This provides us with an opportunity to study the Higgs boson properties, and the Higgs sector in general, at higher energy scales~\cite{offshell_model}. Such a direct probe of Higgs physics at high energies could hold important clues to possible solutions of the naturalness problem of the electroweak scale -- arguably one of the most outstanding problems that has driven the search for new physics at the TeV scale. 

In the absence of new physics signals from extensive searches, especially from the LHC experiments, it is conceivable that the solutions to the naturalness puzzle might have taken a more subtle incarnation, not captured by the usual signatures based on Supersymmetry \cite{susy} (SUSY) or strong dynamics of Composite Higgs \cite{strong}. In this Letter, we adopt a simple illustrative example of such a scenario in which the new physics responsible for partially addressing the little hierarchy problem emerges in the study of Higgs properties at higher energies. We utilize a well-motivated scenario of a new scalar field ($S$) coupled to the Standard Model (SM) Higgs doublet ($H$) through a renormalizable interaction $|S|^2 |H|^2$~\cite{inv_portal}. For appropriate values of the portal sector coupling, such an interaction term can cancel the quadratic divergence to the Higgs mass from top quarks at one-loop, thus alleviating the ``little-hierarchy'' problem \cite{Barbieri}. Though the off-shell Higgs probe to such couplings applies for all assignments of the gauge or global quantum numbers of the scalar field, it  constitutes a model-independent probe to a maximally hidden portal sector, in which the scalars are SM gauge singlets, do not mix with the Higgs boson after electroweak symmetry breaking, are stable, and have masses above the threshold for on-shell Higgs decays. Such a singlet Higgs portal can also be responsible for generating a thermal dark matter relic, and can drive a strongly first-order phase transition to realize electroweak baryogenesis \cite{inv_portal,cline,curtin}.

With this simple scenario in view, we point out for the first time that the presence of such a scalar field leads to measurable deviations in the differential rates for off-shell Higgs production, especially at energy scales  above the $2m_S$ threshold, the amount of deviation from the SM prediction being in accordance with the quadratic divergence cancellation to the Higgs mass. Such deviations arise from the dominant momentum-dependent part of the Higgs self-energy corrections. By studying the gauge-invariant subset of one-loop electroweak corrections from the singlet sector to the process ${p p \rightarrow h^* \rightarrow ZZ}$, we shall demonstrate that it is possible to probe interesting regions of parameter space relevant to the solution of the naturalness problem at the LHC. Thus, the high precision achievable in determining the rate and differential distributions for off-shell Higgs production in the four lepton channel at the high-luminosity phase of the LHC presents us with an excellent opportunity in this regard~\cite{width}.

It was pointed out in~\cite{Han,Englert:2013tya} that any new scalars with an effective coupling of the form $|S|^2|H|^2$  can be probed through the precision measurement of the total rate for $Zh$ production at future lepton colliders, utilizing the universal shift in on-shell Higgs rates from wave-function renormalization. We note that the on- and off-shell production rates for the Higgs signal at the LHC  scale as 
\begin{equation}
\sigma_{\rm on} \propto { {g^2_i(m_h^2) g^2_f(m_h^2)} \over {m_h\Gamma_h}} \  {\rm and}\ 
\sigma_{\rm off} \propto { {g^2_i(Q^2) g^2_f(Q^2)} \over {Q^2}},
\label{eq:sigma}
\end{equation}
respectively, where $g^2_i(Q^2)$ and  $g^2_f(Q^2)$ represent the couplings at the production and decay vertices evaluated at the scale $Q^2$, and $\Gamma_h$ is the Higgs boson total width \cite{offshell,offshell_ex}. Hence, the model-independent interpretation of an on-shell Higgs measurement in terms of particular coupling shifts requires the precise determination of the Higgs boson width as well, for which a future $e^+e^-$ Higgs factory is essential. On the other hand, not only is the off-shell probe of the momentum-dependent part of one-loop scalar singlet corrections a distinct effect, unlike the interpretation of on-shell rate measurements, the interpretation of  off-shell Higgs measurements at the LHC would not require  knowledge of the Higgs boson width. 

To proceed, we introduce an effective Lagrangian for the above scenario in which the singlet sector does not mix with the Higgs field after electroweak symmetry breaking. This is achieved by imposing a $\mathcal{Z}_2$ symmetry under which the singlet sector is odd, and the SM fields are even. The scalar field does not develop a vacuum expectation value, thus the $\mathcal{Z}_2$ symmetry is not spontaneously broken. This can be satisfied by imposing a suitable relation among the parameters of the scalar potential. Thus, the minimal low-energy effective Lagrangian of the gauge singlet scalar sector reads 
\begin{equation}
\mathcal{L} \supset \partial_\mu S \partial^\mu S^* - \mu^2|S|^2 - \lambda_S |S|^2 |H|^2,
\label{eq:lag}
\end{equation}
where $S$ represents a complex scalar field. We note that even though electroweak symmetry breaking effects would generate a contribution to the singlet mass of order $\lambda_S v^2/2$, with $v=246$ GeV being the vacuum expectation value of the Higgs field, the mass parameter of the singlet field, $m_S^2 = \mu^2+\lambda_S v^2/2$ is arbitrary since $\mu^2$ can be of either sign, as long as $m_S^2$ remains positive.

\vskip 0.1cm
\noindent
{\bf II. Analysis}\\  
The possibility to access the Higgs boson contribution to $ZZ$ production with a far off-shell Higgs attracted a lot of attention since its proposal and measurement during the Run-I LHC~\cite{offshell,offshell_ex,offshell_model}. While due to the small Higgs width, such an effect would normally be sub-leading away from the dominant Higgs pole, the large interference with the continuum background $gg\rightarrow ZZ$ converts it into a sensitive measurement of the Higgs contributions. In fact, about $\mathcal{O}(15\%)$ of the Higgs-induced rate of $ZZ$ production resides in the off-shell kinematic region, with the invariant mass of the four leptons $m_{4\ell}>130~\gev$. 

Remarkably, the large off-shell rate allows us to explore Higgs couplings at different energy scales. We utilize this feature to probe the singlet scalar couplings to the Higgs. As mentioned earlier, while the on-shell Higgs search only presents a constant deviation of the Higgs signal strengths due to wave-function renormalization~\cite{Englert:2013tya}, the off-shell Higgs rate can potentially display a momentum-dependence arising from the one-loop scalar singlet corrections, making this analysis even more compelling for the challenging parameter region  $m_h<2m_S$.  
\begin{figure}[t!]
\vspace{0.6cm}
  \includegraphics[width=.23\textwidth]{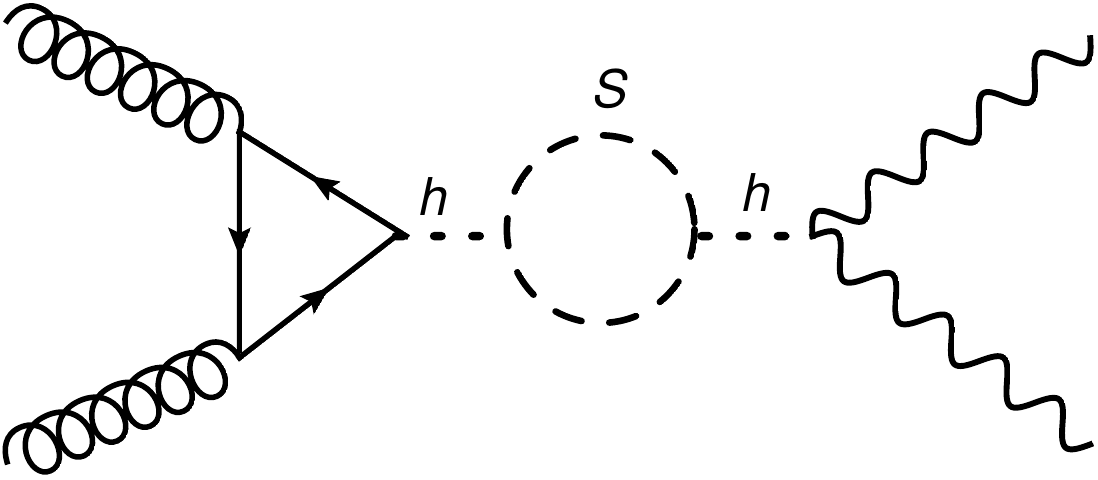}\hspace{0.2cm}
  \includegraphics[width=.23\textwidth]{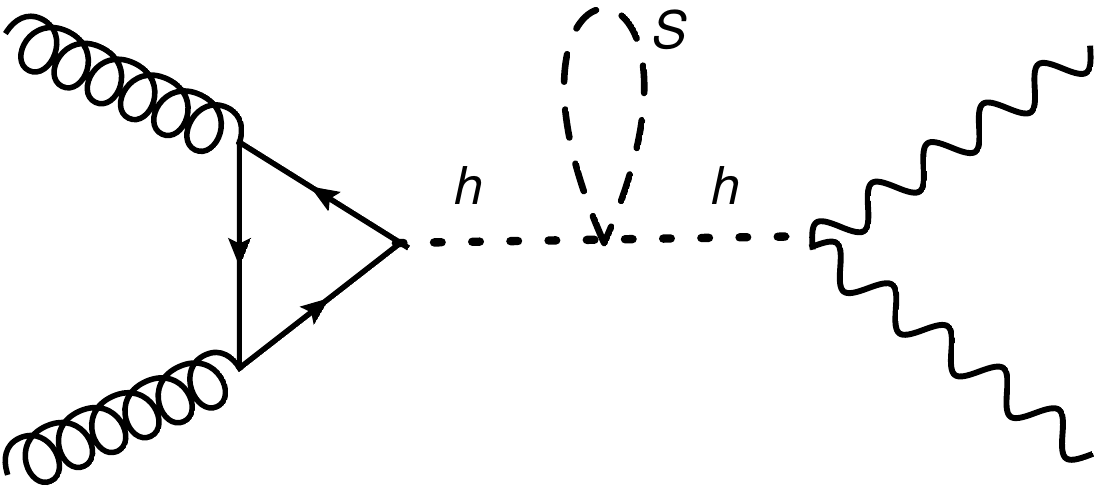}\\
 \vspace{0.3cm}
  \includegraphics[width=.23\textwidth]{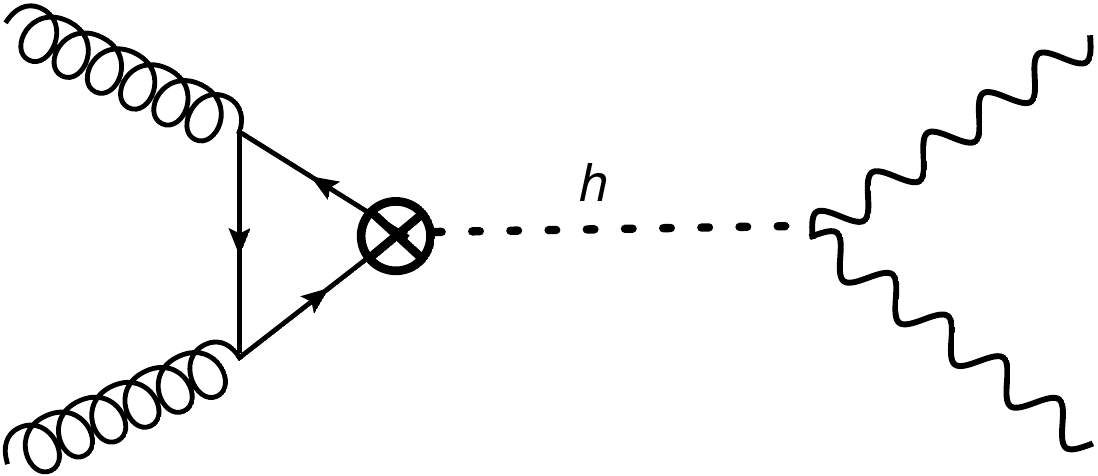}\hspace{0.2cm}
  \includegraphics[width=.23\textwidth]{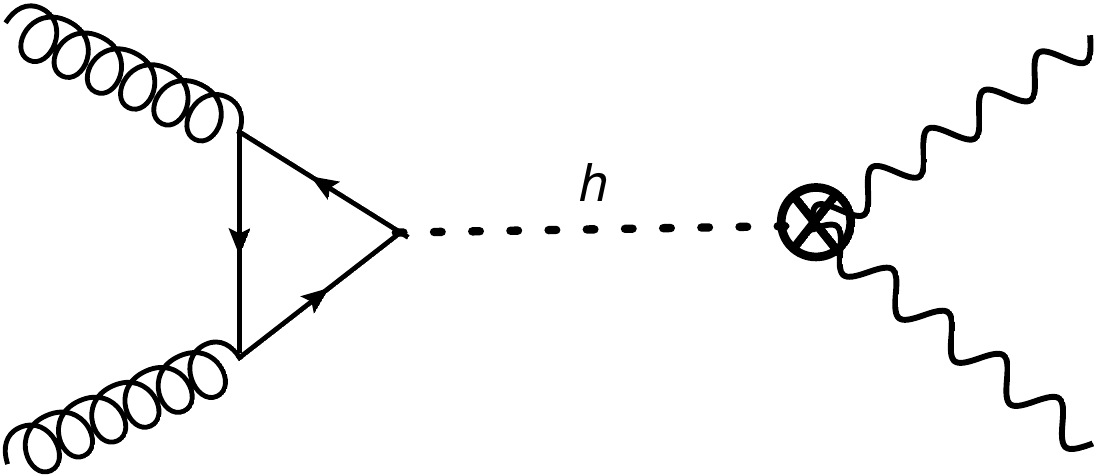}
 \caption{Representative set of Feynman diagrams for the one-loop corrections  to $gg\rightarrow ZZ$ production, in the SM augmented by a
 gauge singlet scalar.}
 \label{fig:feyndiag2}
\end{figure}

\begin{figure}[t!]
  \includegraphics[width=0.4\textwidth]{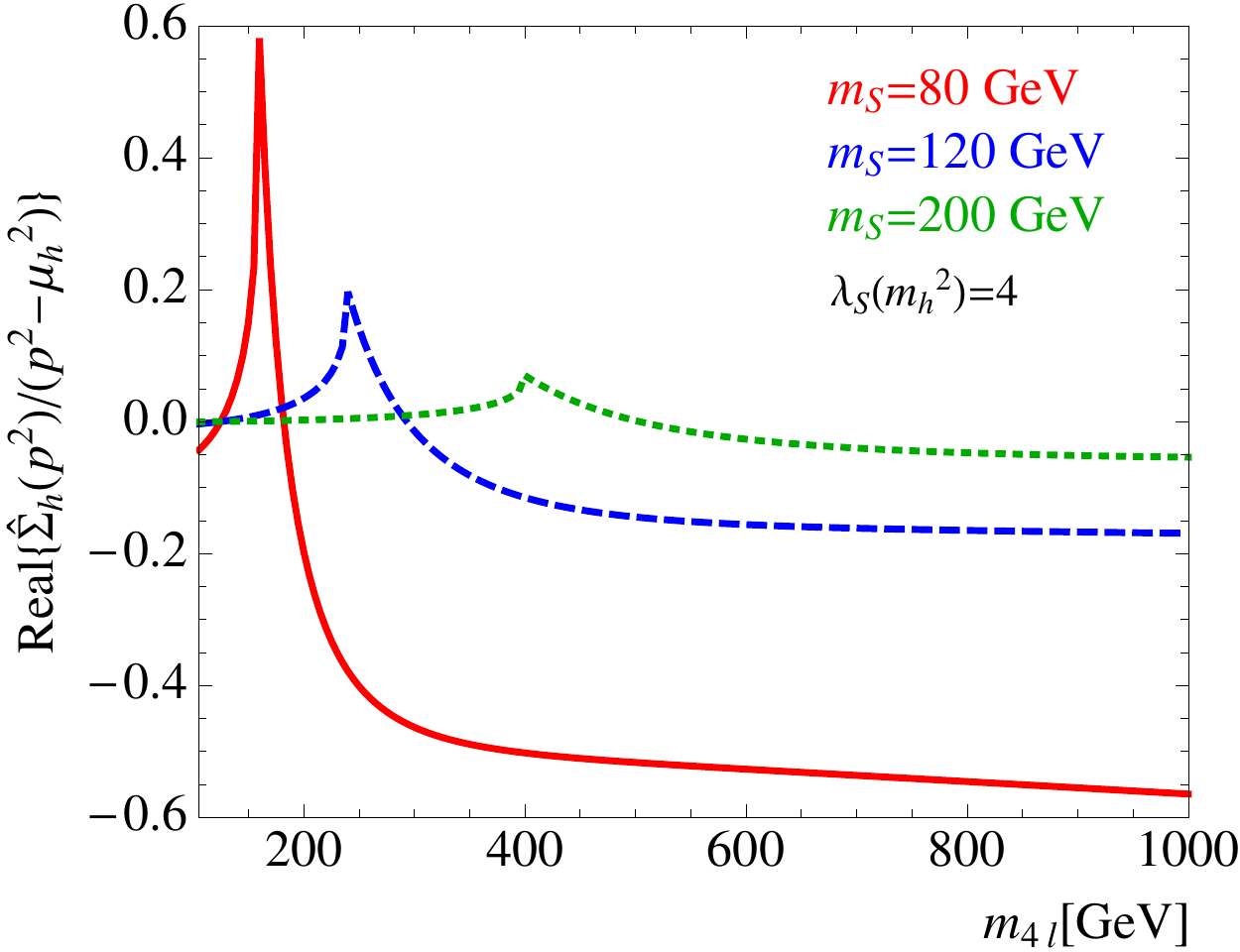}\\
  \vspace{0.3cm}
  \includegraphics[width=0.4\textwidth]{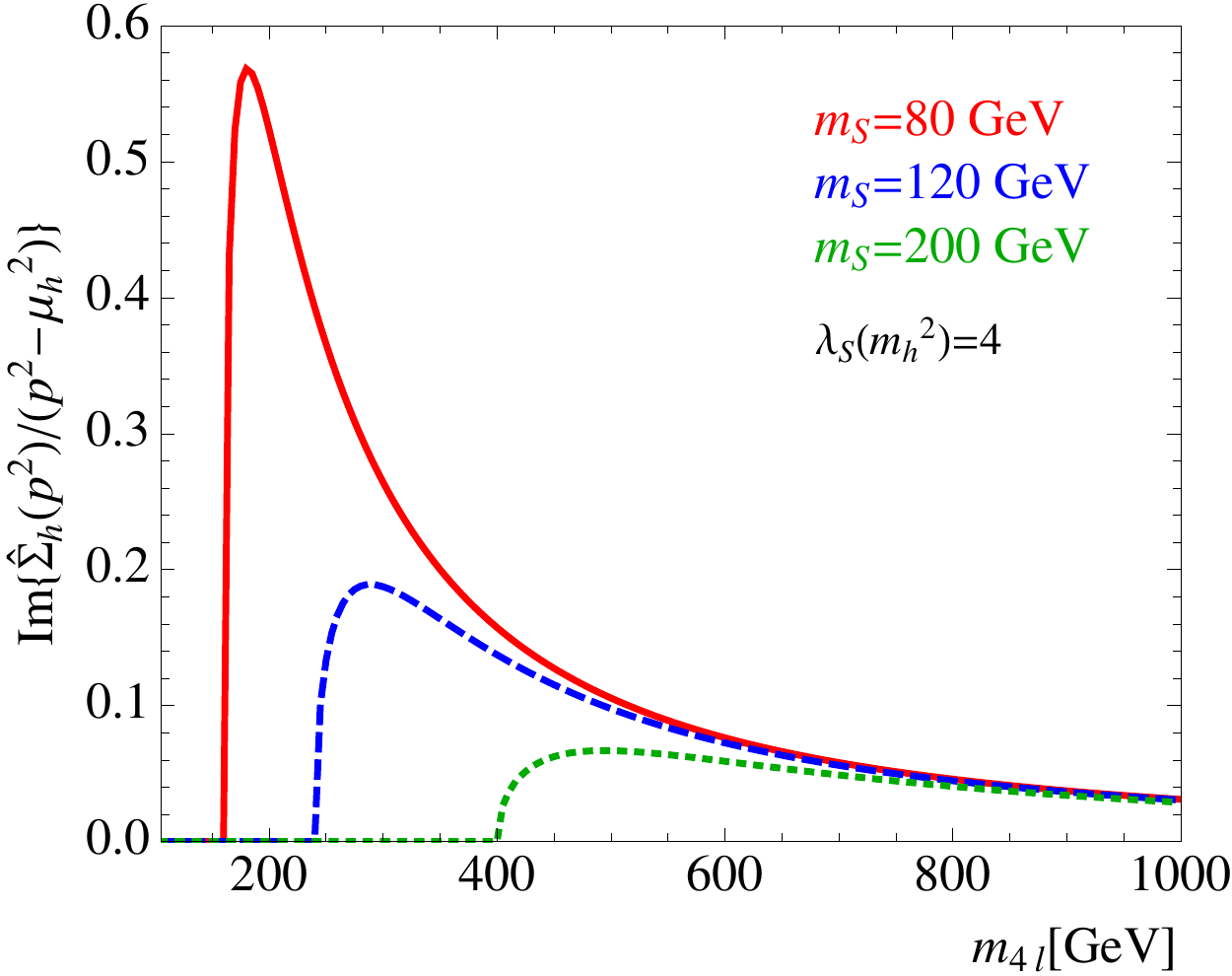}
 \caption{Real (top) and imaginary (bottom) parts of the Higgs boson
 renormalized self-energy corrections $\hat{\Sigma}_h$, scaled by the propagator factor $p^2-\mu_h^2$, as a function of $m_{4\ell}$.}
 \label{fig:selfE}
\end{figure}

\begin{figure}[b!]
  \includegraphics[width=0.42\textwidth]{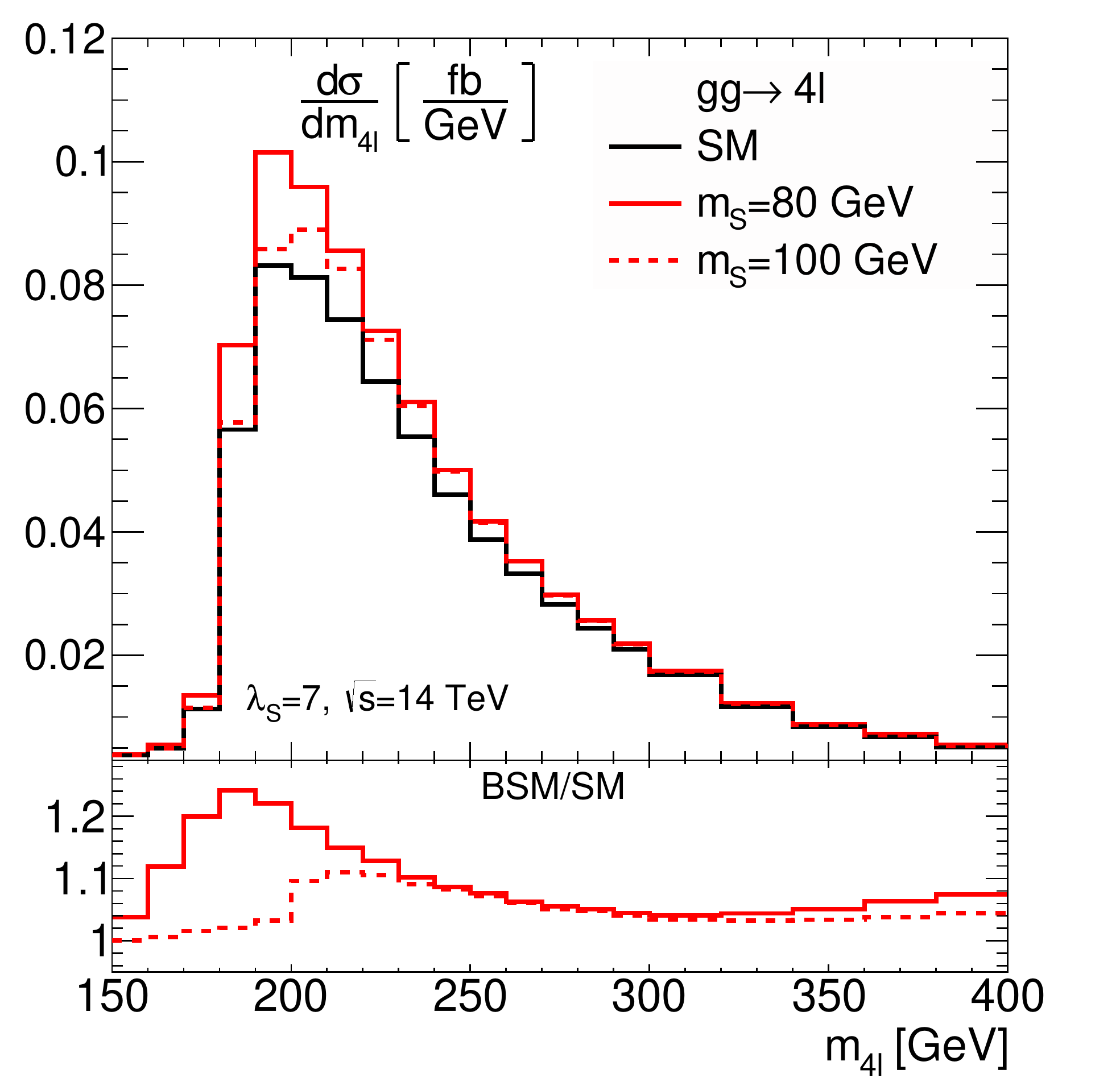}
 \caption{Four lepton invariant mass distribution for the loop-induced $gg\rightarrow 4\ell$ process at the $14$~TeV LHC in the SM (black) and in the presence of an additional scalar gauge singlet (red), 
 including the leading one-loop EW effects from the additional scalar singlet. We display the signal ratio between the scalar singlet model and the SM in the bottom panel.}
 \label{fig:m4l}
\end{figure}
In order to quantify these effects, we consider the ${pp\rightarrow Z^{(*)}Z^{(*)} \rightarrow 4\ell}$ channel at the LHC.
The Higgs boson signal for this channel is generated via gluon fusion with heavy quark loops. There are two major backgrounds to such a process:
$q\bar{q}\rightarrow ZZ$ and ${gg\rightarrow ZZ}$. While the former arises at the tree level, leading to larger rates, the 
latter leads to crucial interference effects with the Higgs signal in the off-shell regime.
We generate the signal and background samples with \textsc{MCFM}~\cite{mcfm}. Spin correlations and off-shell effects
are  fully accounted for, particularly for $Z$-decays to lepton pairs. QCD corrections to the gluon-induced component are
included with an overall K-factor~\cite{offshell}. Using the one-loop scalar integral library \textsc{LoopTools}~\cite{looptools},
we implemented  the scalar gauge singlet corrections, displayed in Fig.~\ref{fig:feyndiag2}, in the \textsc{MCFM} code. 
These represent the leading singlet-induced NLO electroweak corrections, and constitute a separably renormalizable, 
gauge-invariant,  UV-finite subset. Our calculation follows the {\it complex mass scheme}~\cite{complex_mass}, where  the 
renormalized Higgs boson self-energy is defined as
\begin{alignat}{5}
& \hat{\Sigma}_{h}(p^2)=\Sigma_{h}(p^2)-\delta\mu^2_h+(p^2-\mu^2_h)\delta Z_h\;,
\label{eq:complexmass1} 
\end{alignat}
with the complex Higgs mass squared ${\mu_h^2=m_h^2-im_h\Gamma_h}$ and renormalization constants
\begin{equation}
 \delta \mu_h^2=\Sigma_{h}(\mu_h^2)\;, \qquad \delta Z_h=-\frac{d\Sigma_{h}}{dp^2} (\mu_h^2).
\label{eq:complexmass2} 
\end{equation}
We have evolved $\lambda_S(Q^2)$ using the renormalization group equation at one-loop.

It is informative to examine the qualitative features of the scalar singlet one loop contributions.  These are shown in Fig.~\ref{fig:selfE} for the Higgs boson self-energy corrections $\hat{\Sigma}_h$ (scaled by the propagator factor $p^2-\mu_h^2$). One sees the resonant enhancement in the real part and the threshold behaviour in the imaginary part near ${m_h=2m_S}$, that we set out to study next. 

Our search strategy  follows the CMS analysis~\cite{offshell_ex}. The kinematical acceptances are
\begin{alignat}{5}
& p_{T\ell}>10~\gev \;,    & |\eta_\ell| &<2.5 \;,  \notag \\
& m_{4\ell}>150~\gev \;, & m_{\ell \ell'} &>4~\gev \;, \notag \\
& m_{\ell \ell}^{(1)} = [40,120]~\gev \;,  \qquad & m_{\ell \ell}^{(2)} &= [12,120]~\gev \;,
\label{eq:cutflowm4l} 
\end{alignat}
where the last two $m_{\ell \ell}$ refer to the leading and sub-leading opposite charge flavour-matched lepton pair. We use the PDF set \textsc{CTEQ6L1}~\cite{cteq}, and define the factorization and renormalization scales as $\mu_F=\mu_R=m_{4\ell}/2$.\medskip

In Fig.~\ref{fig:m4l}, we present the four-lepton invariant mass distribution for the loop-induced $gg\rightarrow 4\ell$ process in the SM and in the model with an additional scalar gauge singlet at the LHC. We observe that besides shifting the on-shell Higgs rate~\cite{Englert:2013tya}, the higher order corrections to $gg\rightarrow 4\ell$ in the singlet 
model result also in  phenomenologically relevant  kinematic features in the $m_{4\ell}$ distribution, especially above the singlet threshold 
$m_{4\ell}>2m_S$. 

\begin{figure}[t!]
  \includegraphics[width=0.42\textwidth]{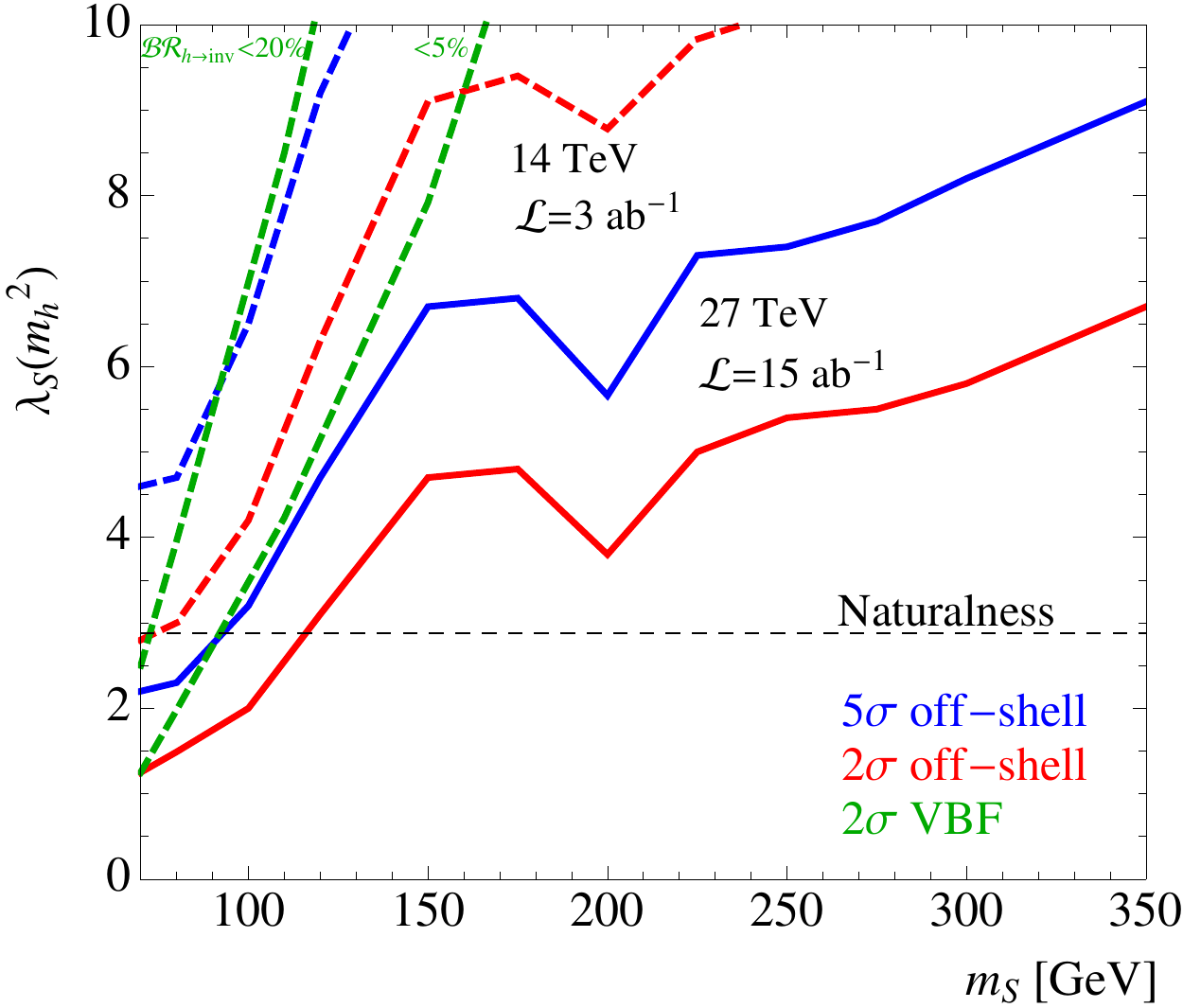}
 \caption{$2\sigma$ (red) and $5\sigma$ (blue) sensitivity on the singlet-Higgs coupling $\lambda_S$  as a function of the singlet mass $m_S$ 
 from the off-shell Higgs analysis at the 14~TeV LHC, with  $\mathcal{L}=3$~ab$^{-1}$  (dashed) and at the 27~TeV LHC, with $\mathcal{L}=15$~ab$^{-1}$ (solid). The $q\bar{q} \rightarrow ZZ$ and $gg \rightarrow ZZ$ backgrounds are taken into account for this estimate.
  For comparison, we also show the reach from 
 VBF production of Higgs above its threshold, assuming the high-luminosity LHC $2\sigma$ level bounds ${\mathcal{BR}(h\rightarrow \text{invisible})<20\%}$  (green dotted) and  $5\%$  (green dashed).}
 \label{fig:bound}
\end{figure}

To quantify the observability of these contributions, we have used a binned log-likelihood analysis on $m_{4\ell}$. The results are shown in Fig.~\ref{fig:bound} for the $2\sigma$  and $5\sigma$ signal sensitivity on $\lambda_S$ evaluated at the scale $m_h^2$ as a function of the singlet scalar mass $m_S$ at the 14~TeV LHC with an integrated luminosity of  $\mathcal{L}=3$~ab$^{-1}$, and at the proposed 27~TeV upgraded LHC with $\mathcal{L}=15$~ab$^{-1}$. Systematic uncertainties have not been included in our analysis, which we expect to be under further control from foreseeable theoretical improvements~\cite{higher_order}. It is observed that there is an enhancement of sensitivity of the off-shell channel for values of $m_S$ close to $m_t$. This is because of the opening of two different thresholds close to each other, namely, the $2m_t$ threshold in the triangle and box diagrams for $ZZ^*$ production, and the $2m_S$ threshold in the radiative correction from the scalar singlet to the same process.

Apart from the search channel proposed in this study, the scalar singlets can also be produced above the Higgs threshold in, for example, vector-boson fusion (VBF), and looked for in the jets and missing momentum final state~\cite{WBF}. We can easily estimate the $2\sigma$ reach for this channel, translating the projected upper bound on the invisible branching ratio for on-shell Higgs in the VBF process, ${\sigma_{VBF}(h)\mathcal{BR}(h\rightarrow \text{invisible})=\sigma_{VBF}(h^{*}\rightarrow SS)}$. We show the VBF search mode reach  in Fig.~\ref{fig:bound}, assuming two high-luminosity $14$ TeV LHC upper bounds of ${\mathcal{BR}(h\rightarrow \text{invisible})<20\%}$ and $5\%$ at $2\sigma$ level~\cite{CMS:2017cwx}. The former refers to a realistic projection of the systematics, and the latter represents an idealistic limit. On comparison with the realistic projection in the VBF channel, we observe that the off-shell Higgs analysis proposed in the present study leads to a better sensitivity on $\lambda_S$ in almost the entire singlet mass range of interest,  $m_S>m_h/2$.  The large interference between signal and background $gg\rightarrow ZZ$, and the longitudinal $Z$ boson enhancement at high energies, guarantees  a larger sensitivity for $pp\rightarrow h^{*}\rightarrow ZZ$, going beyond the usual Breit-Wigner suppression.

\vskip 0.1cm
\noindent
{\bf III. Discussion}\\  
As mentioned earlier, the interpretation of the on-shell Higgs signal strength in terms of specific coupling shifts requires a precise measurement of the Higgs boson total width as well, leading us to rely on a future $e^+e^-$ collider~\cite{Englert:2013tya}. Thus, for example, a modification of the on-shell Higgs signal strengths at the LHC due to the presence of the singlet-Higgs coupling can be compensated for by a simultaneous but independent modification on $\Gamma_h$. In this context, we note that the current uncertainties in the Higgs signal strength measurements in various channels, especially in the bottom pair final state~\cite{ATLAS_bb}, allows for a significant downward shift of the total $\Gamma_h$ compared to its SM value. In order not to induce extra model dependency from assumptions about the Higgs boson width, and to capture the momentum dependence from the new physics, we entirely focus on the off-shell region in this study, with $m_{4\ell}>150$ GeV.

Apart from the information of the scalar particle mass scale, the magnitude of the deviation in the differential rate gives us a measure of the Higgs portal coupling strength. Additionally, the sign of the deviation with respect to the SM prediction for the $M_{ZZ}$ distribution (correlated with the increase or decrease of the destructive interference with the SM $gg \rightarrow ZZ$ contribution)  is also related to the spin of the particle coupling through the Higgs portal $-$ scalar and fermion couplings lead to shifts in opposite directions due to the difference in signs of their self-energy corrections.

We emphasize that any signal observation would have immediate implication for fine-tuning and the possible new physics scale. Assume that the theory is UV completed at a scale $\Lambda$, with new heavy states of this mass scale directly or indirectly (for example, at higher orders in perturbation theory) coupled to the Higgs boson.
The one-loop corrections to the high-scale Higgs mass $M_h$, including the top quark and the singlet scalar contributions, is given by
\begin{eqnarray}
\delta M_h^2  &=& \frac{1}{16 \pi^2} (\lambda_S - 2 N_c y_t^2) \Lambda^2 + \frac{6 N_c y_t^2}{16 \pi^2}  m_t^2 \log \frac{\Lambda^2}{m_t^2}  \nonumber \\
                      & & -\frac{1}{16 \pi^2} \left(\lambda_S m_S^2+\lambda_S^2 v^2\right)\log \frac{\Lambda^2}{m_S^2} ,
\end{eqnarray} 
where $y_t$ is the top quark Yukawa coupling in the SM and the number of colours $N_c=3$.
For the high-scale parameter relation $\lambda_S(\Lambda^2) = 6 y_t^2(\Lambda^2)$,
 the quadratic divergent contribution to the Higgs boson mass from the top quark loop is cancelled exactly by the opposite-sign contribution from the scalar singlet loop. In well-known natural theories like SUSY, such a relation is enforced by symmetry, leading to the cancellation of contributions from the top quark and the scalar top loops. This is 
 indicated by the horizontal dashed line in Fig.~\ref{fig:bound}, with the boundary condition for $\lambda_S$ determined at $\Lambda=10$~TeV, a choice motivated to address the little-hierarchy problem~\cite{Barbieri}. In such a case, the one-loop Higgs mass correction is only logarithmically dependent on the new physics scale and a discovery of a light scalar in this context would allow to relax the new physics scale beyond $100$~TeV if we could tolerate a $5\%$ fine-tuning. In case the new scalar field does not lead to a perfect cancellation, the quadratically divergent piece reappears. The fine-tuning could nevertheless be improved with respect to the SM situation. For example, for around $5\%$ tuning, with $\lambda_S (m_h^2)=4$, we can push the cut-off scale from $2$~TeV to $5$~TeV. We have focussed here on cancelling the dominant quadratic sensitivity in the SM from top quark loops. By a straightforward generalization, the subdominant terms from gauge and Higgs boson loops can also be cancelled -- either with additional new particles, or by  adjusting the singlet-Higgs coupling.

The discovery of such a scalar would also have implications for physics behind some of the pressing problems at the interface of particle physics and cosmology.  A scalar singlet of the weak-scale mass with a large coupling to the Higgs can serve as a component of the total dark matter (DM) density~\cite{inv_portal,cline}. Even though the Higgs boson coupling implies a large spin-independent scattering cross-section in direct detection experiments, the event rate would be small, making such a scenario still consistent with experiments. This is simply because a large annihilation rate in the early Universe implies a small number density surviving after thermal freeze-out. Therefore, the collider probe in off-shell Higgs presented above could be one of the best hopes of detecting such DM particles.

Singlet scalars coupling to the Higgs are also known to be helpful in generating a strongly first-order phase transition during the electroweak symmetry breaking epoch in the early Universe, thus supplying the out-of-equilibrium condition necessary for baryogenesis. It has been demonstrated in previous studies that often such a scenario requires a large coupling of the scalar singlets to the Higgs boson~\cite{curtin}. Hence, if physics beyond the SM can lead to the generation of the required CP-violation as well, a correlated study of the mass and coupling of such singlets could provide circumstantial evidence for successful electroweak baryogenesis.

\vskip 0.1cm
\noindent
{\bf IV. Summary}\\  
We explored the prospects of studying the Higgs sector as a function of the energy scale to probe solutions to the naturalness problem of the Higgs mass. We illustrate this idea with a minimal low energy effective theory that captures the cancellation of quadratic sensitivty of the Higgs mass to high-scale new physics through the top sector, as expected in a natural theory. The model parametrization used, employing only SM singlets, is also minimally observable, thus highlighting the importance of the probe proposed in this study. We have utilized the off-shell Higgs production at the LHC, approaching the new threshold, focusing on the process ${pp \to h^* \to ZZ}$, leading to a clean four-lepton signal. We found that the dominant momentum-dependent part of the one-loop singlet scalar corrections, especially above the new threshold at $2m_S$, leads to a measurable deviation in the differential distribution of the $Z$-pair invariant mass, in accordance with the quadratic divergence cancellation to the Higgs mass. Such a measurement can probe the region of parameter space relevant to the naturalness problem, reaching a 
$2\sigma$ sensitivity at the high-luminosity run of the LHC, and $5\sigma$ at the proposed $27$~TeV LHC upgrade.
With our encouraging results on the off-shell Higgs production, channels other than $ZZ\to 4\ell$ should be systematically considered in the hope to improve the search for a solution of naturalness at the TeV scale.

\textbf{Acknowledgment:}
We thank David Lopez-Val, Brian Batell, David McKeen, Tae Min Hong, Michelangelo Mangano and Lian-Tao Wang for helpful discussions. This work was supported by the U.S.~Department of Energy under grant No.~DE-FG02- 95ER40896 and by the PITT PACC.
DG is also supported by the U.S.~National Science Foundation under the grant PHY-1519175. 
  

\end{document}

%% file: declare.tex
\newcommand{\arxiv}[1]{\href{http://arxiv.org/abs/#1}{arXiv:#1}}

\newcommand\one{\leavevmode\hbox{\small1\normalsize\kern-.33em1}}

\newcommand{\gev}{{\ensuremath\rm GeV}}

\def\slashchar#1{\setbox0=\hbox{$#1$}           
   \dimen0=\wd0                                 
   \setbox1=\hbox{/} \dimen1=\wd1               
   \ifdim\dimen0>\dimen1                        
      \rlap{\hbox to \dimen0{\hfil/\hfil}}      
      #1                                        
   \else                                        
      \rlap{\hbox to \dimen1{\hfil$#1$\hfil}}   
      /                                         
   \fi}

\newcommand{\be}{\begin{eqnarray*}}
\newcommand{\ee}{\end{eqnarray*}}

\newcommand{\bee}{\begin{eqnarray}}
\newcommand{\eee}{\end{eqnarray}}
\newcommand{\beeq}{\begin{equation}}
\newcommand{\eeeq}{\end{equation}}

